\documentclass[]{interact}
\usepackage{epstopdf}

\usepackage[caption=false]{subfig}%

\usepackage[square,numbers]{natbib}
\renewcommand\bibfont{\fontsize{10}{12}\selectfont}

\theoremstyle{plain}
\usepackage{tikz,graphics,color,fullpage,float,epsf}

\theoremstyle{definition}

\usepackage{amsmath,amsfonts,amsthm,bm}
\theoremstyle{remark}

\usepackage{color}
\usepackage{enumitem}

\def\bSig\mathbf{\Sigma}

\usepackage{hyperref}

\begin{document}
\articletype{Application Note}

\title{\textcolor{black}{Network-based Topic Structure Visualization}}

\author{
\name{Yeseul Jeon \textsuperscript{a,b}, Jina Park \textsuperscript{a,b}, Ick Hoon Jin \textsuperscript{a,b} and Dongjun Chung\textsuperscript{c}\thanks{CONTACT Ick Hoon Jin (ijin@yonsei.c.kr) and Dongjun Chung (chung.911@osu.edu)}}
\affil{\textsuperscript{a} Department of Statistics and Data Science, Yonsei University, Seoul, South Korea; \textsuperscript{b}Department of Applied Statistics, Yonsei University, Seoul, South Korea;
\textsuperscript{c}Department of Biomedical Informatics, The Ohio State University, Columbus, Ohio, U.S.A}
}

\maketitle

\begin{abstract}
\textcolor{black}{In the real world, many topics are inter-correlated, making it challenging to investigate their structure and relationships. Understanding the interplay between topics and their relevance can provide valuable insights for researchers, guiding their studies and informing the direction of research. In this paper, we utilize the topic-words distribution, obtained from topic models, as item-response data to model the structure of topics using a latent space item response model. By estimating the latent positions of topics based on their distances toward words, we can capture the underlying topic structure and reveal their relationships. Visualizing the latent positions of topics in Euclidean space allows for an intuitive understanding of their proximity and associations. We interpret relationships among topics by characterizing each topic based on representative words selected using a newly proposed scoring scheme. Additionally, we assess the maturity of topics by tracking their latent positions using different word sets, providing insights into the robustness of topics. To demonstrate the effectiveness of our approach, we analyze the topic composition of COVID-19 studies during the early stage of its emergence using biomedical literature in the PubMed database. The software and data used in this paper are publicly available at \url{https://github.com/jeon9677/gViz}.}
\end{abstract}


\section{Introduction}
Most real world data rarely consists of independent topics but rather multiple topics that are often inter-correlated. By exploring the relationships and dependencies among topics, researchers can gain a deeper understanding of the underlying structure and dynamics of their research domain. This understanding can help identify emerging trends, uncover research gaps, and highlight areas of high importance or potential impact. It can help researchers be knowledgeable of research areas that are emerging and/or significant to the field and allow them to put their efforts on important ones. There have been attempts to model correlated structure among topics by modeling the correlated structure within the topic model or combining the topic model with statistical models. For instance, \citet{blei2007correlated} regarded the topic of correlation as a structure of heterogeneity. To capture the heterogeneity of topics, they suggested the correlated topic model (CTM), which models topic proportions to exhibit correlation through the logistic normal distribution within Latent Dirichlet Allocation (LDA). \citet{rusch2013model} combined LDA with decision trees to interpret the topic relationships from Afghanistan war logs. It classified the topics using tree structures, which helped to understand the different circumstances in the Afghanistan war. However, these approaches do not directly allow us to investigate the degree of correlation and closeness among topics. To address this, \citet{sievert2014ldavis} attempted to quantify the relationships and proximity between topics by calculating the distances between topics. Specifically, they used multi-dimensional scaling (MDS) to embed the topics on a metric space based on a distance matrix calculated from the output of a topic model (LDA). However, MDS requires specification of similarity and dissimilarity measures to construct the distance matrix and the results can be significantly affected by the choice of measures in spite of their ad hocness. In practice, it is often difficult to determine the appropriate measures for calculating and defining distances between topics. Hence, it is beneficial to have a unified approach to estimate the structure of topics, which allows us to avoid such ad hoc choices and guarantee more consistent results.
One of the statistical methods for handling correlated structure is network analysis. Specifically, network model estimates the relationships among nodes based on their dependency structure. Moreover, it can provide global and local representation of nodes at the same time. In this context, \citet{mei2008topic} tried to identify topical communities by combining a topic model with a social network approach. Specifically, it estimated topic relationships given the known network structure and tried to separate out topics based on connectivity among them. However, in this approach, the network information needs to be provided to construct and visualize the topic network, which is not a trivial task in practice. 

To address this, here we propose an alternative approach to examine the structure of topics using the information from the output of topic models. Specifically, our approach leverages a latent space network model, which estimates the relationships among nodes by estimating topics' latent positions. The latent positions can then be used to quantify the relationships among nodes, i.e., the distances between topics. In our study, we treat the output of the topic-words distribution as a topic-words matrix $\mathbf{X}$, which can be seen as an item-response data \cite{embretson2013item} with an inter-correlated structure. Using this matrix, we simultaneously estimate the relationship between topics and words, which yields the latent position of topics. When two topics share similar word meanings, their latent positions are located close to each other, making it easier to understand relationships between topics based on their associated words. Furthermore, when topics share common words, their latent positions tend to be located near the origin. Conversely, topics that incorporate more distinctive and exclusive words compared to other topics tend to be positioned farther away from the origin. This behavior elucidates the relationship between the words associated with topics and their respective positions in the latent space. This also enables us to visualize the structure of topics using the estimated latent positions.

In this approach, we interpret the relationships among topics by characterizing topic based on its representative words. If two topics are closely located in the latent positions, they might deliver similar meanings in documents. To make the process to select the meaningful words within the topics more efficient and objective, it is desirable to have a score calculated based on the information provided by our analytical framework. Previously \citet{airoldi2016improving} evaluated the topics using the novel score called \textit{FREX}, which quantifies words' closeness within the topic based on the word frequency and exclusivity. Here, $\textit{FREX}_{i,j}$ reflects how word $j$ is exclusively close to topic $i$ compared to other topics. However, relying solely on the frequency of words in documents is not enough to identify key words within topics that reflect the association among topics. To address this, here we propose a score that takes into account both (i) the probability of words belonging to each topic based on the frequency of associated words using the topic model; and (ii) how closely words are exclusively related to each topic by considering the relevance between topics using the network model. By putting these two pieces of information together, our analytical framework can pick up information about relativeness and exclusivity. 

It is also important to acknowledge the fact that topics also vary in the sense of how mature it is and whether it is common or specific. We can evaluate this by checking whether latent positions of topics are robust to different word sizes. This is based on the rationale that mature topics and/or those consisting of common words might remain similar and stable upon changes in word sets, compared to emerging or specific topics. Hence, to understand the changes in latent positions of topics, we track the latent positions of topics by differing word sets. To select words sets of different sizes, we used two criteria to select informative words: high coefficient variation and high maximum probability from topic-words distribution. First, words with large variations in probabilities among topics are expected to be more important because if a word has low variation across topics, it is likely that the word does not represent any topic specifically. We used coefficient of variation to characterize words' dispersion among topics. Second, a meaningful word should have a high probability in at least one topic. Even when a word has high variation among topics, if that word have only low probabilities across topics, it might not be informative to differentiate topics.

The rest of this article is organized as follows. In Section 2, we introduce our dataset and the Gaussian version of the latent space item response model \cite{Jeon:2021}, which estimates latent positions of topics based on their latent distances between words. In addition, we introduce our scoring scheme that helps to select representative words from each topic to enhance the understanding of the topic structure. In Section 3, we apply our approach to the COVID-19 literature to evaluate and demonstrate the usefulness of our approach. In Section 4, we summarize our topic structure visualization with conclusion.

\section{Methodology}

\subsection{Data}
Since the emergence of the worldwide pandemic of COVID-19, relevant research has been published at a dazzling pace, which yields an abundant amount of big data in biomedical literature. Due to the high volume of relevant literature, it is practically impossible to follow up the research manually. Furthermore, in the early stages of research when a specific topic has not yet been established, many studies are likely to be connected to each other. Therefore, by examining which topics are being discussed together, we can more easily understand the research direction of COVID-19. For instance, if some researchers want to investigate a specific topic, say `COVID-19 symptoms', we want to answer the questions like the following: (1) Which set of words are associated with `COVID-19 symptoms'? (2) Are there any other topics related to `COVID-19 symptoms', which can be used for extending and elaborating research? (3) Is `COVID-19 symptoms' a common or specific topic? 

To model the COVID-19 literature, we downloaded the COVID-19 articles published from the PubMed database (\url{https://pubmed.ncbi.nlm.nih.gov}) with a time-frame between December 1st, 2019, and August 3rd, 2020, coinciding with the date of the WHO's designation of COVID-19 as a pandemic. We collected articles whose titles contain ``coronavirus2'', ``covid-19'' or ``SARS-CoV-2'', which resulted in a total of 35,585 articles. Since COVID-19 was a worldwide pandemic that spread at an unprecedented rate, some articles that examine COVID-19 have been published with only an abstract. After eliminating articles without abstracts (i.e., only titles or abstract keywords), our final text data contained a total of 15,015 abstracts.

Based on the abstracts, we constructed the corpus using the abstract keywords that concisely captured the messages delivered by the paper. To enrich the corpus, we further used the {\tt word2vec} approach \cite{mikolov2013efficient} to train against relationships between nouns from the abstract and the abstract keywords. Specifically, {\tt word2vec} extracted nouns from abstracts, which were embedded near the abstract keywords, and added those selected words to the corpus. Using the trained {\tt word2vec} network with 256 dimensions, we selected ten words from the abstract nouns that were near each abstract keyword. We provide the details of our training strategy for the {\tt word2vec} network in the Supplemental Materials (see Appendix A). The corpus construction resulted in 9,643 words from 15,015 documents. We further filtered out noise words, including single alphabets, numbers, and other words that are not meaningful, e.g., `p.001', `p.05', `n1427', `l.', and `ie'. Finally, to obtain more meaningful topics, we removed common words like `data', `analysis', `fact', and `disease'. We provide the full list of filtered keywords in the Supplemental Materials (see Appendix B).

To obtain the topic-words matrix $\mathbf{X}$, we implemented the topic modeling. Since the abstracts of the COVID-19 literature are categorized as short text, conventional topic models including LDA often suffer from poor performance due to the sparsity problem. \citet{yan2013biterm} made important progress in the modeling of short text data, the so-called Biterm Topic Model (BTM). Unlike the LDA, BTM replaced words with bi-terms, where a bi-term is defined as a set of two words occurring in the same document. This approach attempted to compensate for the lack of words in a short text by pairing two words, thereby creating more words in the document. This is based on observation that if two words are mentioned together, they are more likely to belong to the same topic. Using the BTM, we were able to obtain a topic-words distribution, which is essentially a matrix of topics and their associated words $\mathbf{X}$. This matrix $\mathbf{X}$ was then used as an input for the latent space item response model.
 
\subsection{Estimating Topic Relationships Using Latent Space Item Response Model} 

\citet{hoff2002latent} proposed the latent space model, which expresses a relationship between actors of a network in an unobserved ``social space'', so-called latent space. Inspired by \citet{hoff2002latent}, \citet{Jeon:2021} proposed the latent space item response model (LSIRM) that viewed item response as a bipartite network and estimated the relationship between respondents and items using the latent space. In order to achieve our objective of estimating the topic structure and visualizing their relationships, we utilized the concept of latent positions. These latent positions allow us to estimate the relevance of topics based on the distances between topics and associated words. By measuring the closeness between topics in terms of their latent positions, we can intuitively represent their distances. To estimate the latent positions of topics, we employed LSIRM with a bipartite network representation using the matrix $\mathbf{X}$. In this network, each item represents a topic $i$ and respondents correspond to word $j$. However, the original LSIRM proposed by \citet{Jeon:2021} cannot be directly applicable here because it was designed for binary item response dataset, where each cell in the item response data has a binary value (0 or 1). On the contrary, here our input data $\mathbf{X}$ has continuous probabilities indicating how likely each word belongs to each topic. Therefore, in order to apply LSIRM to our input data $\mathbf{X}$, we use Gaussian version of LSIRM, which is described in detail below. The modified Gaussian version of LSIRM can be written as  
\begin{align*}
    \begin{split}
        x_{i,j} \mid \boldsymbol{\Theta} &= \boldsymbol\beta_i +\boldsymbol\theta_j - || {\bf v}_i - {\bf u}_j || + \epsilon_{i,j}, \\ 
        \epsilon_{i,j} &\sim \mbox{N} (0, \sigma^2),
    \end{split}
    \label{eq:LSIRM}
\end{align*}
where $x_{i,j}$ indicates the log scale probability that word $j$ belongs to topic $i$, for $i = 1, \cdots, P$ and $j = 1, \cdots, N$. Because the original LSIRM use logit link function to handle the binary data, here we use the linearity assumption between $x_{i,j}$ and the attribute part with the relevance part. We added an error term $\epsilon_{j,i} \sim N(0, \sigma^2 )$ to satisfy the normality equation. We use the notation $\boldsymbol{\Theta} = \{\boldsymbol{\beta}=\{\boldsymbol\beta_i\}, \boldsymbol{\theta}=\{\boldsymbol\theta_j\}, \bf{U}= \{{\bf u}_j\},  \bf{V}=\{{\bf v}_i\} \}$, and $|| {\bf v}_i-{\bf u}_j ||$ represents the Euclidean distance between latent positions of word $j$ and topic $i$. We estimate the relevance between topics by measuring the distances between topics. Note that the distances between topics are measured based on their relationships with words.  Specifically, if two topics share similar word sets, we consider them to have a close relationship and estimate their relationship accordingly. Given the model described above, we use Bayesian inference to estimate parameters in the Gaussian version of LSIRM. We specify prior distributions for the parameters as follows:
\begin{align*}
    \boldsymbol\beta_i &\sim \text{N}(0,1), \\
    \boldsymbol\theta_j \lvert \sigma^2 &\sim \text{N}(0,\sigma_{\theta}^2),\quad \sigma^2 >0 \\\sigma^2 &\sim \text{Inv-Gamma}(a , b),\quad a_>0 , \quad b>0 \\
    \sigma_{\theta}^2 &\sim \text{Inv-Gamma}(a_\sigma , b_\sigma),\quad a_\sigma>0 , \quad b_\sigma>0 \\
    \bf{u}_j &\sim \text{MVN}_d (\mathbf{0, I}_d) \\
    \bf{v}_i &\sim \text{MVN}_d (\mathbf{0, I}_d),
\end{align*}
where $\mathbf{0}$ is a length-$d$ vector of zeros and $\mathbf{I}_d$ is the $d \times d$ identify matrix. We treat $\beta_i$ as a fixed effect and $\theta_j$ as a random effect in order to account for the variability among words. By including a random effect, the model acknowledges that different words may exhibit unique patterns and behaviors towards specific topics. The posterior distribution of LSIRM is 
\begin{equation*}
    \pi(\boldsymbol{\Theta}, \sigma^2 |\bf{X} ) \propto \prod_{j}\prod_{i} \mathbb{P} \left (x_{i,j}| \boldsymbol{\Theta} \right) 
    \prod_j \pi(\boldsymbol\theta_j | \sigma_{\theta}^2 ) \pi(\sigma_{\theta}^2) \prod_i \pi(\boldsymbol\beta_i )  
    \prod_j \pi(\bf{u} _j )  \prod_i \pi( \bf{v}_i)  \pi(\sigma^2)
\end{equation*}
and we use Markov Chain Montel Carlo (MCMC) to estimate the parameters of LSIRM. In this way, we can obtain latent positions of $\mathbf{u}_j$ and $\mathbf{v}_i$ on the $\mathbf{R}^{d}$. Since we are interested in constructing the topic network, we utilize $\mathbf{v}_i, i=1,\cdots, P$ and make it as matrices of ${\bf A} \in \mathbf{R}^{P\times d}$ where row indicates $P$ number of topics and column indicates $d$ number of dimension of coordinates. 

The latent positions of topics are assumed to follow a standard normal distribution. As a result, there is a tendency for these latent positions to spread out from the origin. Specifically, when topics share common words, their latent positions tend to be located near the origin, indicating their similarity. On the other hand, topics with more distinctive and exclusive words compared to other topics tend to be positioned farther away from the origin, indicating their uniqueness. This behavior illustrates the relationship between the words associated with topics and their corresponding positions in the latent space. Therefore, latent positions of topics not only enable us to visualize the dependency structure of topics, but also help understand the topic structure through their spreading tendency.

Here we further improve the interpretation of relationships among topics by tracing how topics' latent positions change as a function of word sets. In other words, we compare topics' latent positions ${\bf A}_k$ from the various sets of matrices $\mathbf{X}_k, k=1,\cdots,K$ to estimate and trace their relevance among topics based on word sets with different sizes. After proceeding with the LSIRM model with various sets of matrices $\mathbf{X}_k$, we can obtain matrices of ${\bf A}_k$, composed of coordinates of each topic. Since the distances between pairs of words and topics, which are crucial for inferring latent positions, can be subject to variations due to rotation, translation, and reflection, multiple possible realizations of latent positions can exist. These variations in the likelihood function necessitate careful consideration to ensure accurate estimation and inference \cite{hoff2002latent, shortreed2006positional}. In order to tackle this invariance property for determining latent positions, we implemented within-matrix Procrustes matching \cite{borg2005modern} as post-processing of MCMC samples. Specifically, we implemented a Procrustes matching two times. First, we implement so-called within-matrix matching within the MCMC samples for each topic's latent positions generated from LSIRM. 
Second, we implemented so-called between-matrix matching for the estimated matrices to locate topics in the same quadrant. To align the latent positions of topics, we need to set up the baseline matrix, denoted as ${\bf A}_\text{max}$, which maximizes the dependency structure among topics. To measure the degree of dependency structure, we take the average of the distances of topics' latent positions from the origin. The longer distance of latent positions from the origin implies a stronger dependency on the network. It helps nicely show the change of topics' latent positions because those rotated positions ${\bf A}_k$ from each matrix ${\bf X}_k$ are based on the most stretched-out network from the origin. As a result, we can obtain the re-positioned matrices $\mathbf{A}^*_k$, which still maintain the dependency structure among topics but are located in the same quadrant.

With the oblique rotation, the interpretation of axes can be further improved and topics can be categorized based on these axes. For this purpose, we applied the {\tt oblim} rotation \cite{jennrich2002simple} to the estimated topic position matrix $\mathbf{A}^*_k$, using the R package {\tt GPAroation} \cite{bernaards2005gradient}. We denote the rotated topic position metric by $\mathbf{B}_k$. To interpret the trajectory plot showing traces of topics' latent positions, we extracted the rotation information matrix ($\mathbf{R}$) resulting from an oblique rotation as the baseline matrix $\mathbf{B}_\text{base}$. Then, we multiplied each matrix ($\mathbf{B}_k$) by the rotation matrix ($\mathbf{R}$) to plot the topics' latent positions. 

\subsection{Scoring the Words Relation to Topics}

By combining the idea from \citet{airoldi2016improving} with the relevance information among topics in our problem, here we propose the score $s_{i,j}$ which measures the exclusiveness of word $j$ in the topic $i$. We define score $s_{i,j}$ as 
\begin{equation}
    \begin{split}\label{eq:score}
        s_{i,j} &= \left( \frac{w_1}{2-\text{ECDF}_{\boldsymbol\delta_{.,j}}(\delta_{i,j})} \right)
        +\left(\frac{w_2}{2-\text{ECDF}_{\boldsymbol\delta_{i,.}}(\delta_{i,j})} \right)\\
        &+\left(\frac{w_3}{1+\text{ECDF}_{\boldsymbol\gamma_{.,j}}(\gamma_{i,j})} \right)
        +\left(\frac{w_4}{1+\text{ECDF}_{\boldsymbol\gamma_{i,.}}(\gamma_{i,j})} \right) ,
    \end{split}
\end{equation}
where $w_1,w_2,w_3,$ and $w_4$ are the weights for exclusivity (here, we set $w_1=w_2=w_3=w_4=0.25$) and $\text{ECDF}$ is the empirical CDF function. Here, $\delta_{i,j}$ denotes the probability of word $j$ belonging to topic $i$ given by BTM, and $\gamma_{i,j}$ is the distance between the latent position of word $j$ and topic $i$ estimated by LSIRM. The higher value of $\delta_{i,j}$ indicates the closer relationship between word $j$ and topic $i$ and the smaller value of $\gamma_{i,j}$ indicates the shorter distance between word $j$ and topic $i$. To make the meaning of the shorter distance and the higher probability consistent as both contribute to the higher scores, first, we subtracted $\text{ECDF}_{\boldsymbol\delta_{.,j}}(\delta_{i,j})$ and $\text{ECDF}_{\boldsymbol\delta_{i,.}}(\delta_{i,j})$ from 2. Note that here we use 2 to avoid the zero in the denominator. For example, if $\delta_{i,j}$ has a higher probability within the topic $i$ and between the other topics and word $j$, then both $2-\text{ECDF}_{\boldsymbol\delta_{.,j}(\delta_{i,j})}$ and $2-\text{ECDF}_{\boldsymbol\delta_{i,.}(\delta_{i,j})}$ will have smaller values. This means that the word $j$ is distinctive enough to represent the meaning of topic $i$. Second, if the latent distance between word $j$ and topic $i$ has the shorter distance within topic $i$ than between the other topics and word $j$, then both $1+\text{ECDF}_{\boldsymbol\gamma_{.,j}(\gamma_{i,j})}$ and $1+\text{ECDF}_{\boldsymbol\gamma_{i,.}(\gamma_{i,j})}$ have the smaller values contributing to a higher score $s_{i,j}$. Here we added 1 to each denominator term, preventing the denominator from becoming zero. Based on $s_{i,j}$, we can determine whether the word $j$ and the topic $i$ are close enough to be mentioned in the same document. By collecting the high-score words, we can characterize the topics.

\section{Application}

\subsection{Topic-words matrix of COVID-19 Biomedical literature using BTM} 

To implement BTM, we set the topic number to 20. For the hyper-parameters, we assigned $\alpha$ = 3 and $\beta$ = 0.01. Since our main goal is to estimate the topic structure, we empirically searched and determined the hyper-parameters of BTM. We explain the details of the modeling the topic model in the Supplemental Materials (see Appendix C). The posterior distribution of the topic-words was estimated using the Gibbs sampler. Specifically, we generated samples with 50,000 iterations after the 20,000 burn-in iterations and then implemented thinning for every $100^{\text{th}}$ iteration. 

In each topic-words distribution obtained from BTM, words with high probabilities characterize the topic. Since the topic-words distribution contains all the word sets from the corpus, using all word sets can lead to redundancy. To confirm this, we draw histogram of log-transformed probabilities and it shows bimodal topic-words distributions. This pattern indicates that there are some words that had low probabilities of belonging to a specific topic, whereas the mode in the center corresponds to the words that have probabilities high enough to characterize the meaning of the topic. Therefore, it might be more desirable to estimate topic structure using only the words corresponding to the mode in the center rather than all the words. On the other hand, our exploratory analysis indicated that more than 1,000 words are needed to represent topics properly. Based on this rationale, we decided to use at least 1,000 words to estimate the positions of topics in the latent space based on the positions of words. Given the selected minimum number of words, we extracted meaningful words that can characterize topics based on the coefficient of variation and maximum probability.

Given the coefficient of variation and maximum probability, we obtained multiple matrices corresponding to the top 60\% to 40\% of words determined based on the two criteria. The numbers of words corresponding to the $60^{\text{th}}$ and $40^{\text{th}}$ percentiles were 2,648 and 1,095, respectively. We used 21 sets of matrices, $\mathbf{X}_k (k = 40\%, \cdots, 60\%)$, as the LSIRM input data, where their dimensions were ranged from $2,648 \times 20$ to $1,095 \times 20$.
In this way, we obtained the 21 sets of matrices ${\bf X}_k$ and we considered 20 topics for all the matrix sets. We applied a log transform to the values in these matrices, which originally ranged between 0 and 1, to convert them into continuous values ranging from $-\infty$ to $\infty$. MCMC was implemented to estimate topics' latent positions $\bf{V}= \{{\bf v }_i\}$ where $i=1,\cdots,20$. The MCMC ran 55,000 iterations, and the first 5,000 iterations were discarded as burned-in processes. Then, from the remaining 50,000 iterations, we collected 10,000 samples using a thinning of 5. To visualize relationships among topics, we used two-dimensional Euclidean space. Additionally, we set 0.28 for $\boldsymbol\beta$ jumping rule, 1 for $\boldsymbol\theta$ jumping rule, and 0.06 for ${\bf w}_j$ and ${\bf z}_i$ jumping rules. Here, we fixed prior $\boldsymbol\beta$ follow $N(0,1)$. We set $a_{\sigma}=b_{\sigma}=0.001$. 

LSIRM takes each matrix ${\bf X}_k$ as input and provides the ${\bf A}_k$ matrix as output after the Procrustes-matching within the model. Since we calculated topics' distance on the 2-dimensional Euclidean space, ${\bf A}_k$ is of dimension $20 \times 2$. We visualized the topic structure using the baseline matrix ${\bf A}_{\max}$ so that we can compare topics' latent positions without having identifiable issues from the invariance property. From ${\bf A}_k$, we calculated the distance between the origin and each topic's coordinates. The closer distance of a topic position from the origin indicates the weaker dependency with other topics. In our interrogation of ${\bf A}_k$, we found that the dependency structure among topics starts to be built up from ${\bf A}_{47\%}$. There are two possibilities that can lead to low dependency. First, it is possible that a small number of words could distinguish the characteristics of their topics from the other topics. Second, it is possible that most of the words were commonly shared with other topics. We provide the distance plot of ${\bf A}_k$ in the Supplemental Materials (see Appendix D). Based on this rationale, we chose ${\bf A}_{47\%}$ as the baseline matrix ${\bf A}_{\max}$. With this baseline matrix ${\bf A}_{47\%}$, we implemented the Procrustes matching to align the direction of the topic's latent positions from each matrix ${\bf A}_k$. Using this process, we could obtain the ${\bf A}_k^{*}$ matrix matched to the baseline matrix ${\bf A}_{47\%}$. We named the identified topics based on top ranking words using the ${\bf A}_{47\%}$ matrix because the baseline matrix ${\bf A}_{47\%}$ has the most substantial dependency structure. 
We applied {\tt oblimin} rotation to the estimated topic position matrix $\mathbf{A}_{k\%}^{*}$ using the R package {\tt GPArotation}\cite{bernaards2005gradient}, and obtained matrix $\mathbf{B}_k$ $k=40\%, \cdots, 60\%$ with the rotation matrix $\mathbf{R}$. By rotating the original latent space, we could obtain more interpretable axes for the identified latent space (e.g., determining the meaning of a topic's transition based on the X-axis or Y-axis).

\begin{figure*}[ht!]
   \subfloat[Topics' latent positions of ${\mathbf B}_{47\%}$  \label{baseline}]{%
      \includegraphics[width=0.3\textwidth]{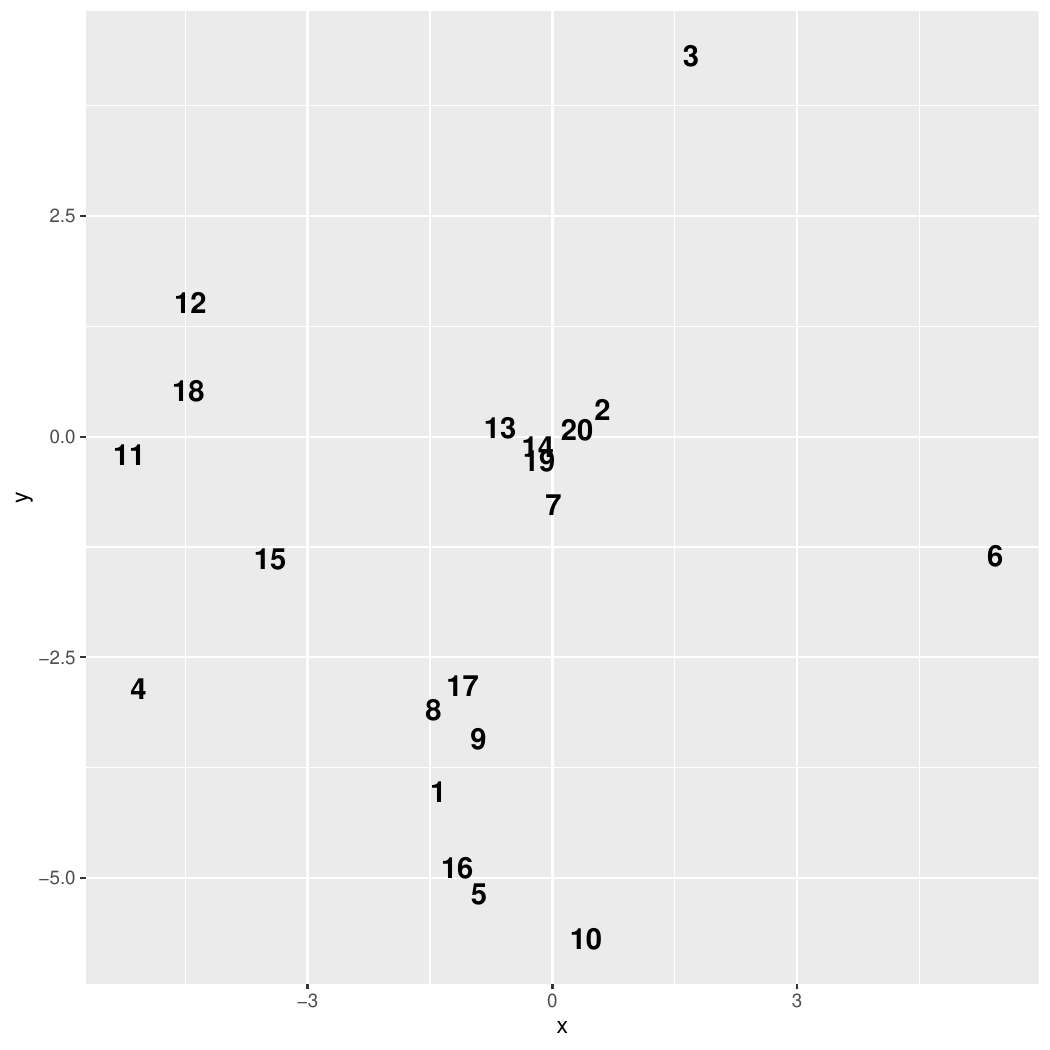}}
\hspace{\fill}
   \subfloat[The score and affinity plot of Topic 19 in ${\bf B}_{47\%}$ \label{commontopic} ]{%
      \includegraphics[width=0.3\textwidth]{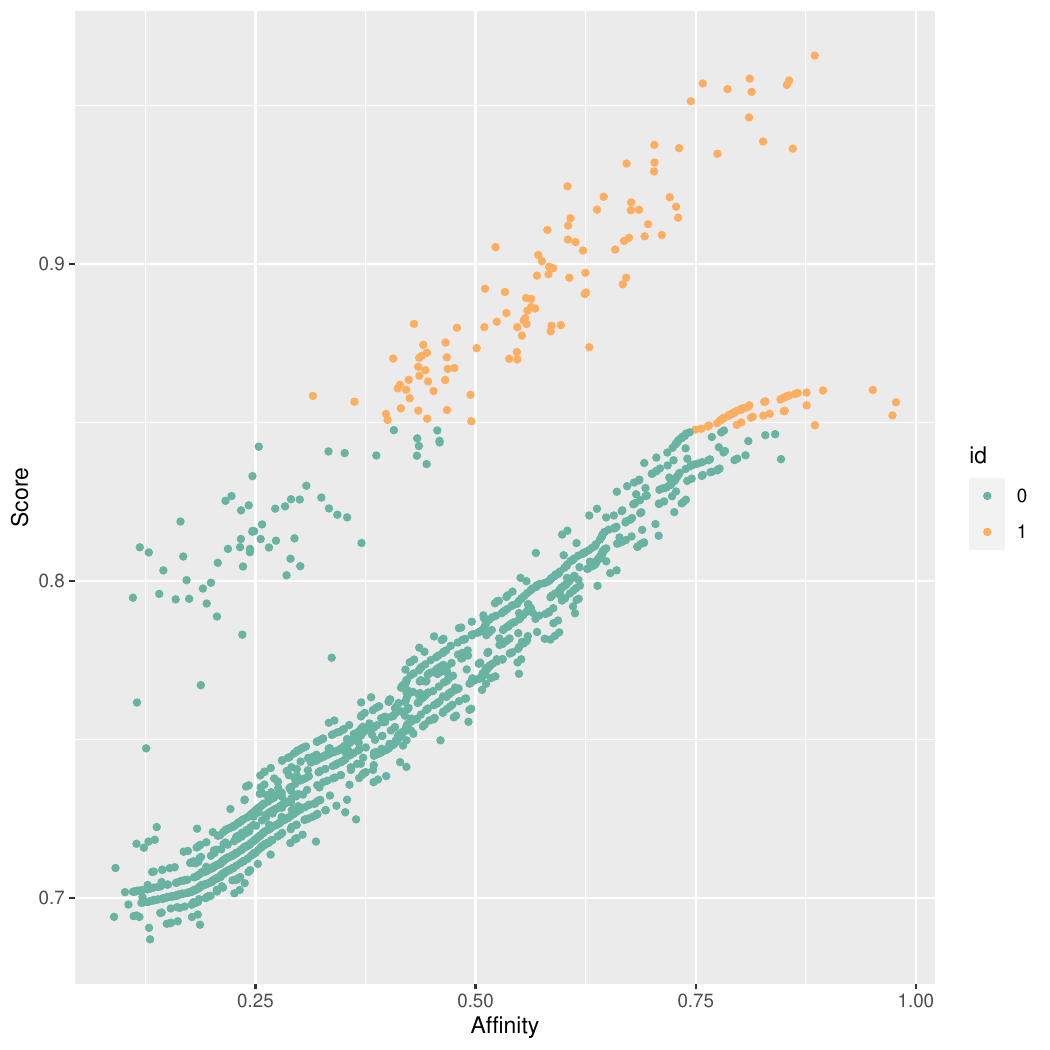}}
\hspace{\fill}
   \subfloat[The score and affinity plot of Topic 3 in ${\bf B}_{47\%}$ \label{uniquetopic}]{%
      \includegraphics[width=0.3\textwidth]{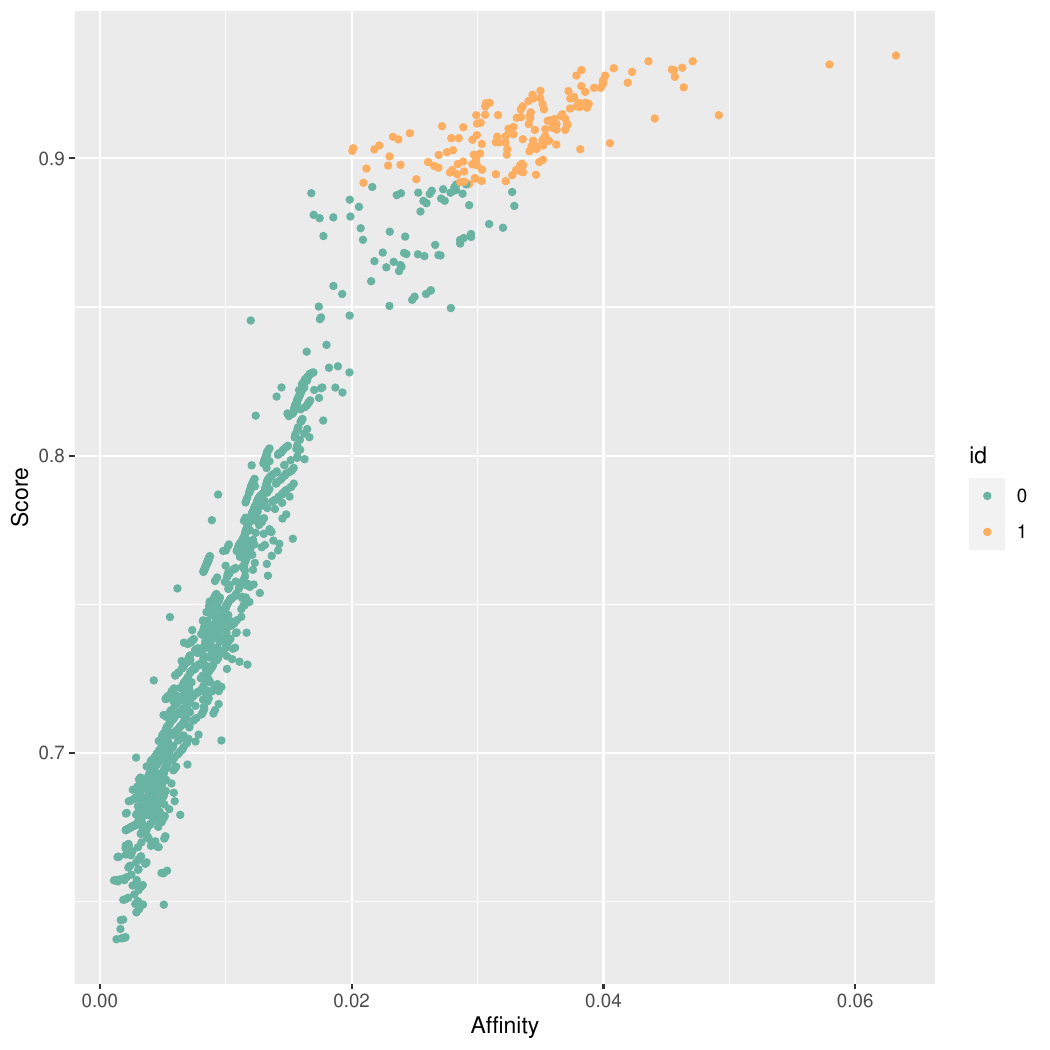}}\\
\caption{\label{scoreplotcompare}If topics are located in the center in latent position (a), two distinct subgroups are observed in the score and affinity plot (b). Otherwise, such subgroup patterns are not observed in the score and affinity plot (c). In (b) and (c), orange color (id of $1$) indicates top 20\% of words of high score words while green color (id of $0$) indicates the remaining words.}
\end{figure*}

\subsection{Words selection for each topic based on the score} 

We calculated the score ${s}_{i,j}$ for each word $j$ with topic $i$ based on each $\bf{B}_{k}$ for $k=40\%, \cdots, 60\%$. Figure~\ref{scoreplotcompare} shows the two different scenarios of the plot of $s_{i,j}$ and affinity ($\exp{(-\gamma_{i,j})}$) in ${\bf B}_{47\%}$; one with a high score and high affinity, and the other with a low score but high affinity. Figure~\ref{baseline} shows the estimated latent positions of topics obtained from ${\bf B}_{47\%}$. We can see that topics 2, 7, 13, 14, 19, and 20 are located around the origin, indicating their general nature and similarity in terms of their word composition. These common topics share many words in common with other nearby topics, making it challenging to discern their unique characteristics solely based on word probabilities (Figure \ref{commontopic}). On the other hand, the remaining topics are positioned away from the origin, indicating that they have distinctive characteristics compared to other topics (Figure \ref{uniquetopic}). For these topics, it is important to understand their direction and identify which topics are closely related. We provide score and affinity plot of each topic from ${\bf B}_{47\%}$ in the Supplemental Materials (see Appendix E).
We extracted the top 20\% of words with high values in score to interpret each topic and we collected the meaningful words that commonly appear on the top of the list among every portion of words set ${\bf B}_{k}, k=40\%, \cdots, 60\%$. Table~\ref{topicname} shows the interpretation of each topic made based on the selected words with its top score words. We provide more detailed interpretation of the topics in the Supplemental Materials (see Appendix F).

\begin{table}[tt]
\centering
\caption{\label{topicname} \textbf{Interpretation of the topic based on $\mathbf{A}^{*}_{47\%}$ matrix.} The abbreviation is marked by an asterisk and the more detailed interpretation can be found in the Supplemental Materials (see Appendix G).}
\resizebox{\columnwidth}{!}{%
\begin{tabular}{lllll}
  \hline
Topic  & Name & Top score words  \\
  \hline
1 & Lung Scan & subpleural  & crazy paving   & bronchogram\\ 
 & & (0.963) & (0.957) & (0.950)\\
2 & Compound and Drug & Protein Data Bank &  papain-like & intermolecular \\
& & (0.938) & (0.936) &  (0.934) \\ 
3 & Bacteria, Nano, and Diet & $\text{DHA}^{\ast}$ & biofilm  & vancomycin-resistant  \\
&& (0.968) & (0.968) & (0.967) \\ 
4 & Treatment of Other Diseases & sarcoma &  arthroplasty & neoadjuvant \\
&& (0.948) & (0.944) & (0.943)  \\
5 & Symptoms (Cardiovascular) & vein & antithrombotic & $\text{VTE}^{\ast}$  \\
&& (0.906) & (0.899) & (0.898)  \\ 
6 & Molecular-level Response to Infection & $\text{NLRP3}^{\ast}$ & metalloproteinase & upregulation  \\
&& (0.929) &(0.927) &(0.926) \\ 
7 & COVID-19 Risk Prediction Markers & alanin & $\text{BUN}^{\ast}$ & prealbumin \\
&& (0.976) & (0.972) &(0.969)\\ 
8 & Literature Review & Prospero DB &  Wanfang DB  & meta-analys  \\
 && (0.967) & (0.966) & (0.944)\\
9 & Symptom and Comorbidity & petechiae & guillain-barre syndrome & ageusia\\
&&  (0.958) &  (0.955) & (0.952) \\ 
10 & Cardiovascular & infarction & thromboprophylaxis & $\text{VTE}^{\ast}$ \\
&& (0.912) &(0.909) &  (0.909) \\  
11 & Social Impact & classroom  & pedagogy& zoom  \\
&&  (0.925) & (0.911)  &(0.906) \\ 
12 & Financial and Economical Impact & macroeconomics & monetary & profit \\
&& (0.929) &(0.928) & (0.928) \\ 
13 & Statistical Modeling & Weibull& least-square & $\text{MAE}^{\ast}$\\ 
&& (0.984) & (0.975) & (0.966) \\
14 & COVID-19 Test & $\text{LFIAs}^{\ast}$  & transcription-$\text{PCR}^{\ast}$ & Wantai \\
&&  (0.979) & (0.974) &(0.972) \\ 
15 & Psychological/ Mental Issues & anxious&$\text{PTSD}^{\ast}$& post-traumatic \\
&& (0.942) & (0.940) &  (0.935)\\
16 & Lung Scan &  procalcitonin & ground-glass & opacity \\
&& (0.924) &  (0.922) &(0.916) \\
17 & Treatment & reintubation &  multi-centric& Hydroxychloroquine \\
&& (0.943) & (0.943) &  (0.937) \\
18 & Air, Mask, Breathing & laryngoscope& facepiece & supraglottic \\
&& (0.978) & (0.978) & (0.975)\\ 
19 & Prevention of COVID-19 & alcohol-based & decontamination &rub\\
&& (0.944) &  (0.941) & (0.919)\\ 
20 & Immunology and Virus &  $\text{MHC}^{\ast}$ &  multi-epitope & immunodominant \\
&& (0.976) & (0.975) &(0.973)  \\ 
\hline
\end{tabular} %
}
\end{table}

\subsection{COVID-19 Topic Structure Visualization} 

\begin{figure}[htbp]
\centering
\includegraphics[width = 0.9\linewidth]{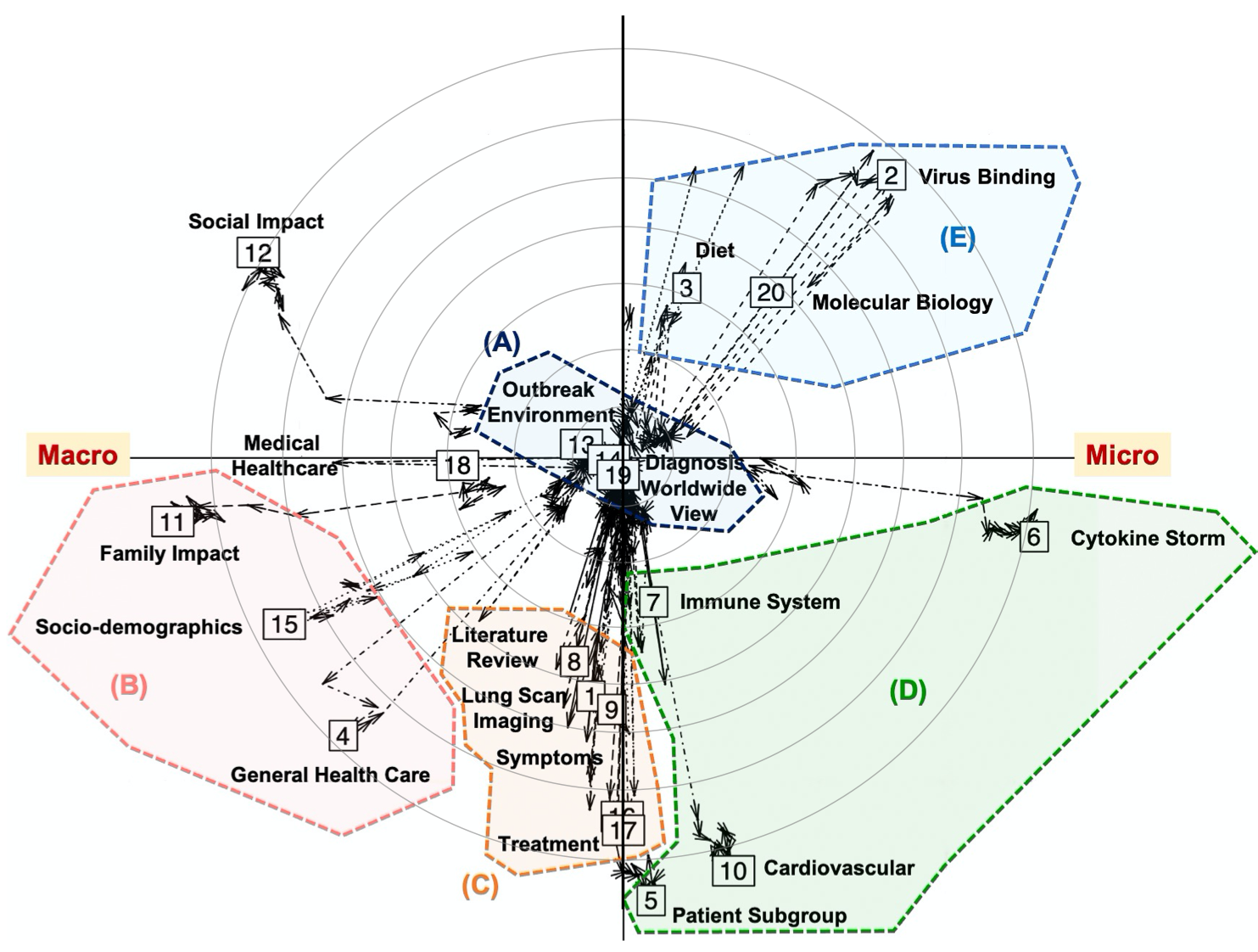}
\caption[]{Topic architecture visualization of COVID-19 literature with three topic clusters (A) - (C). The arrow indicates the direction of latent positions by increasing the words size. (A) COVID-19 and its impacts on different areas. (B) Symptoms and treatment of COVID-19 and relevant studies.  (C) Molecular studies associated with the COVID-19 infection.}
\label{finalresults}
\end{figure}

We interrogated what kinds of topics have been extensively studied in the biomedical literature on COVID-19. In addition, we also studied how those topics were related to each other, based on their closeness in the sense of latent positions. We also partially clustered the topics based on their relationships using the quadrants. This allows us to check which studies about COVID-19 are relevant to each other and could be integrated. Figure \ref{finalresults} displays the trajectory plot and it shows how topics were positioned on the latent space and how these topics make the transition across word sets of different sizes. Here the direction of arrows refers to how topics' coordinates changed as a function of the numbers of words, where each arrow moved from ${\bf B}_{60\%}$ to ${\bf B}_{40\%}$ as the number of words decreased. 

According to Figure~\ref{finalresults}, we observed two distinct groups. The first group consists of topics that have common words shared among them, while the second group consists of topics that have their own distinct words. First, the topics `Statistical Modeling', `COVID-19 Test', and `Preventing COVID-19' (Topics 13, 14, and 19, respectively) were located in the center of the plot (the former group). This indicates that no matter how many words were used to estimate the topics' latent positions, those topics remained as general topics and shared many words with other topics. This makes sense given the fact that the outbreak of COVID-19 and the testing of COVID-19 were mentioned in a large proportion of literature, and statistical models were key tools to study the COVID-19 pandemic. For example, among the 137K publications mentioning ``COVID-19'' in the PubMed database, more than 76,000, 64,782 ,and 44,602 publications also mentioned ``outbreak'', ``testing'', and ``prevention'' respectively. Second, the topics `Social Impact (financial and economical)', `Social Impact', `Cardiovascular', and `Cytokine Storm' (Topics 12, 11, 10, and 6, respectively) were located away from the center in the plot (the latter group), which implies their dependency structures with other topics. These topics usually stay on the boundary of the plot regardless of the number of words because they mainly consist of unique words. Finally, the topics like `Psychological Mental Issues', `Relevance of Treatment', and `Compound and Drug' (Topics 15, 4, and 2, respectively) start from the origin, move away from the origin for a while, and then return to the origin. This implies that it could not maintain the nature of the topic when fewer words were considered, and it is likely that those topics are either ongoing research or burgeoning topics that have not been studied enough yet.

Next, we interpreted the topics' meaning based on their latent positions, specifically using subsets of topics divided by directions. Since we implemented the oblique rotation that maximizes each axis' distinct meaning, we can render meaning to a direction. Figure~\ref{finalresults} indicates that there are three topic clusters. First, the center cluster denoted as (A) in Figure~\ref{finalresults} are about the outbreak of COVID-19 and its impacts on different areas, including the outbreak of COVID-19, diagnosis or testing of COVID-19 using statistical models, the impacts on general social, financial, or economic problems, and the mental pain of facing the pandemic shock of COVID-19. Second, the topics located at the bottom of the plot (cluster (B) in Figure~\ref{finalresults}) are related to the symptoms of COVID-19 and their treatment. For instance, there are studies of COVID-19 related to their symptoms, such as cardiovascular diseases and lung scan images. On the other side, there are topics related to treatment and risk prediction markers for COVID-19. These subjects pertain to  `Literature Review (8)', `Lung Scan Imaging (1)', `Symptoms Comorbidity (9),' `Lung Scan (16)', `Treatment (17)', `Relevance of Treatment (4)', `COVID-19 Risk Prediction Markers (7)', and `Cardiovascular (10)'. Finally, the cluster (C) is related to what happens inside our body in response to COVID-19, e.g., cytokine storm, immune system response to the COVID-19 infection, compounds of drugs, and mechanisms of SARS-CoV-2.

In summary, we identified three main groups of the COVID-19 literature; the outbreak of COVID-19 and its effects on the society, the studies of symptoms and treatment of COVID-19, and the impacts of COVID-19 on our body, including molecular changes caused by COVID-19 infection. We can derive another insight from the locations of clusters. Specifically, from Cluster (A) to Cluster (C), we can observe a counter-clockwise transition from macro perspectives to micro perspectives. Specifically, this flow starts with the center Cluster (A) related to the occurrence of COVID-19 and the social impact of COVID-19, followed by studies of the symptoms and treatment of COVID-19 (Cluster (B)) and then ends with Cluster (C), which are related to micro-level events, e.g., how SARS-CoV-2 binding occurs and how the immune system responds to upon the COVID-19 infection.

\section{Conclusion}
We utilize the latent space item response model to estimate the topic structure in this manuscript. First, by estimating the latent positions of topics, we are able to visualize the topic structure in Euclidean space. Specifically, we apply this approach to the COVID-19 biomedical literature, which helps improve our understanding of COVID-19 knowledge in the biomedical literature by evaluating the networks of topics via their latent positions derived from topic sharing patterns with words. Additionally, visualization of the topic structure provides an intuitive depiction of the relationships among topics. This embedded derivation of topic relationships will reduce the burden on data analysts because it does not require prior knowledge about relationships between topics, e.g., a connectivity matrix or distance matrix. 

Second, we introduce a score, denoted by $s_{i,j}$, that aids in interpreting the topics without requiring expert knowledge. Without such scoring scheme, topics can be misinterpreted due to the inter-correlated structure of some topics, which can make estimated probabilities from topic models inaccurately reflect the exclusivity of words associated with each topic. Moreover, without such scoring scheme, identifying meaningful words for each topic and characterizing the topic using those words often requires significant expert knowledge. Our approach overcomes these challenges by providing a means to determine the exclusiveness of word $j$ with topic $i$ without relying on subjective interpretation.

Third, we interpret the topic structure by tracing their latent positions as a function of different levels of word richness. This feature has two important properties: (1) it could measure the main location of the topic, which is steadily positioned in a similar place in spite of differing network structures; and (2) we could distinguish popular topics mentioned across articles from recently emerging topics by scanning the latent position of each topic. Specifically, if a particular topic shares most of the words with other topics, it is more likely to be located in the center. In contrast, if a specific topic consists mostly of words unique to that topic (e.g., a rare topic or an independent topic containing its own referring words), it is more likely to be located away from the center. For example, in the context of COVID-19, it is more likely that common subjects like `outbreak' and `diagnosis' are located in the center, while more specific subjects like `Cytokine Storm' are located more outside.

The proposed framework can still be further improved in several ways. Although the distance between each topic and relevant words is taken into account in our model when estimating topics' latent positions, simultaneous representation and visualization of words are still not embedded in the current framework. We believe that adopting a variable selection procedure to determine key words can potentially address this issue and this will be an interesting future research avenue.

\section*{Acknowledgements}

This work was supported by the National Institutes of Health [grant numbers R21-HG012482, R21-CA209848, U01-DA045300, R01-GM122078, U54-AG075931 awarded to DC], Yonsei University Research Fund [grant number 2019-22-0210 awarded to IHJ] and the National Research Foundation of Korea [grant number NRF 2020R1A2C1A01009881; Basic Science Research Program awarded to IHJ]. The funders had no role in study design, data collection and analysis, decision to publish, or preparation of the manuscript.

\section*{Disclosure statement}
No potential conflict of interest was reported by the authors.

\bibliography{reference}
\label{lastpage}

\vspace*{-8pt}

\end{document}


\title{Network-based Topic Structure Visualization} 

\author{Yeseul Jeon, Jina Park, Ick Hoon Jin, and Dongjun Chung}
\date{}

\subtitle{\bf Supplemental Materials}

\maketitle

\section{Training {\tt word2vec} Network}

To train the words' network, \citet{mikolov2013distributed} suggested the negative-sampling approach, which fits the network of words by training the near words and unrelated words, and this approach was reported to be efficient in vectorizing the words' relationships. \citet{goldberg2014word2vec} expanded the negative-sampling strategy to model the words and the contexts through the joint modeling of words and contexts, which makes the problem non-convex. 

Before implementing the negative sampling, we need to assign the window size that defines how many neighbors of words to consider. For example, let's assume that we want to select four neighboring words of `princess'. According to the word embedding network, there are only a few words located close in the latent Euclidean space, e.g., `horse', `money',  `king', `queen', `princess', `prince', `palace', `flowers', and `...'. Since we want to select four neighboring words, we set the window size to 2 so that two words from each side centering `princess' can be considered. Therefore, we define `king', `queen', `prince', and `palace' as near words for the word `princess'. On the other hand, we can sample 20 negative words that are not included in these near word sets. By repeating this process, it trains the words' network that reflects the context. In this way, we could obtain the word embedding network with 256 dimensions.

\section{Filtered Keywords}
\begin{table}[htt]
    \centering
    \caption{Filtered keywords}
    \begin{tabular}{ccccccc} \hline
         p.001 & find & z9.738 & n102 & p0.0001 & p.05 & measure\\ 
         p0.03 & n1427 & sd  & coronaviru & case & number & studi \\ 
         date & result & hope & refer & major & b & n  \\ 
         besid & ie & l. & fact & e.g. & h & p \\ 
         half & virus & viru & disease & coronavirus & data & rate \\ 
         factor & method & test & model & analysis & health & death \\ \hline
    \end{tabular}
    \label{tab:filtered}
\end{table}

\clearpage
\section{Estimating Latent Topics using Biterm Topic Model}
The abstract is an excellent source for understanding the overall text. Since most abstracts are limited to 200 words, they could be regarded as short texts. Therefore, we use Biterm Topic Model (BTM) for literature mining. To implement BTM, we first need a word corpus, and, then, we extract biterms from each paper, which is input for the BTM. We extract information from the text using morphology analysis, one type of natural language processing technique. Specifically, it splits each word with a suffix to identify the base element unit of a term. Among the basic units of words, we first extract nouns in their most basic forms. This set of nouns is called a corpus. We further expand the corpus by adding relevant words using {\tt word2vec}, which vectorizes distances among words in the Euclidean latent space according to their similarity in meaning. We extract neighboring words based on these distances and enlarge the word sets by collecting these words. Following this, it is necessary to understand the overall semantic structure of a given text. It summarizes the latent structures, called topics, by identifying topics and estimating their corresponding clusters of words. That is, we can estimate topics and their distributions of words using topic modeling.

The BTM is based on the following assumptions describing how bi-terms and topics jointly related to each other: 
\begin{itemize}
    \item Two words are assumed to belong to some hidden clustered set of words. 
    \item It is assumed that there are hidden topics, each of which represents similar meaning of words. 
    \item It is assumed that those hidden clustered sets of words are hidden topics. 
\end{itemize}

The likelihood of BTM consists of both topic-word distribution and topic distribution. Therefore, we need two sets of parameters to estimate the topic distribution $\boldsymbol{\theta}$ and the topic-word distribution $\boldsymbol{\phi}_z$. The whole likelihood of BTM is constructed as follows. First, the prior distribution for words ($\boldsymbol{\phi}_{z}$) is set to Dirichlet distribution with hyper-parameter $\beta$ while the prior distribution for topics is set to Dirichlet distribution with hyper-parameter  $\alpha$. Next, we represent the topic with the latent variable $z$, which follows Multinomial distribution with parameter $\theta$. Likewise, each word follows Multinomial distribution with parameter $\boldsymbol{\phi}_z$ so that each word can be generated from a specific topic. Therefore, there are three parameters to estimate, including $\boldsymbol{\phi}_{z}$, $\boldsymbol{\theta}$, and ${z}$.  Using the collapsed Gibbs sampling \citep{liu1994collapsed}, we can construct distribution of $\boldsymbol{\phi}_{w|z}$ and $\mathbf{\theta}_z$ with estimated statistics $n_{w|z}$ and $n_z$, given as follows: 
\begin{equation}\label{eq:3}
    \boldsymbol\phi_{w|z} = \frac{n_{w|z}}{\sum_{w}{n_{w|z}}+M\beta}, \quad \mbox{and} \quad 
    \boldsymbol\theta_{z}=\frac{n+\alpha}{|B|+K\alpha}
\end{equation}

The BTM is based on the following assumptions describing how bi-terms and topics jointly related to each other: 
\begin{itemize}
    \item Two words are assumed to belong to some hidden clustered set of words. 
    \item It is assumed that there are hidden topics, each of which represents similar meaning of words. 
    \item It is assumed that those hidden clustered sets of words are hidden topics. 
\end{itemize}

\noindent
Based on these assumptions, the likelihood of BTM consists of both topic-word distribution and topic distribution. Therefore, we need two sets of parameters to estimate the topic distribution $\boldsymbol{\theta}$ and the topic-word distribution $\boldsymbol{\phi}_z$. The whole likelihood of BTM is constructed as follows. 
First, the prior distribution for words ($\boldsymbol{\phi}_{z}$) is set to Dirichlet distribution with hyper-parameter $\beta$ while the prior distribution for topics is set to Dirichlet distribution with hyper-parameter  $\alpha$. 
Next, we represent the topic with the latent variable $z$, which follows Multinomial distribution with parameter $\theta$. Likewise, each word follows Multinomial distribution with parameter $\boldsymbol{\phi}_z$ so that each word can be generated from a specific topic. Therefore, there are three parameters to estimate, including $\boldsymbol{\phi}_{z}$, $\boldsymbol{\theta}$, and ${z}$.
The likelihood construction process can be summarized as follows:

\begin{enumerate}
     \item Draw a whole topic distribution from $\theta \sim \mbox{Dirichlet}(\alpha)$. 
     \item For each biterm b from biterm set B, calculate to which among latent topics $z$ those two words belong: ${z} \sim \mbox{Multinomial}(\theta)$
     \item Draw a topic-word distribution of topic $z$ from $\boldsymbol{\phi}_z \sim \mbox{Dirichlet}(\beta)$.
     \item Draw probabilities for two words from the topic-word distribution corresponding to the selected topic:  $w_i,w_j \sim \mbox{Multinomial}(\boldsymbol{\phi}_z)$
\end{enumerate}
After the above steps, the joint likelihood of all biterm set B is calculated as: 

\begin{equation}\label{eq:1}
   p(B) = \prod_{i,j}\sum_{z}\boldsymbol\theta_z\boldsymbol\phi_{i|z}\boldsymbol\phi_{j|z}
\end{equation}

The conditional posterior for a latent topic ${z}$, $p({z}|{z}_{-b},\mathbf{B},\alpha,\beta)$, is given as: 
\begin{equation}\label{eq:2}
    p(\mathbf{z}|\mathbf{z}_{-b},\mathbf{B},\alpha,\beta) \propto (n_{z}+\alpha)\frac{(n_{w_i|z}+\beta)(n_{w_j|z}+\beta)}{(\sum_{w}{n_{w|z}}+M\beta)^2}
\end{equation} 
where $n_{z}$ is the number of times that the biterm $b$ is assigned to topic $z$, $z_{-b}$ refers to the topic without biterm $b$, and $n_{w|z}$ is the number of times of the word $w$ assigned to the topic $z$. Because a direct application of the Gibbs sampling for $p(z|z_{-b}, \mathbf{B},\alpha,\beta)$ sometimes can result in lack of convergence due to dependency between variables, we use the collapsed Gibbs sampling to constrain unnecessary parameters by integrating out \citep{liu1994collapsed}. In particular, because the prior distributions are Dirichlet distributions, $\boldsymbol{\phi}_{z}$ and $\mathbf{\theta}$ can be integrated out. 
After some iterations, we can construct distribution of $\boldsymbol{\phi}_{w|z}$ and $\mathbf{\theta}_z$ with estimated statistics $n_{w|z}$ and $n_z$, given as follows: 
\begin{equation}\label{eq:13}
    \boldsymbol\phi_{w|z} = \frac{n_{w|z}}{\sum_{w}{n_{w|z}}+M\beta}, \quad \mbox{and} \quad 
    \boldsymbol\theta_{z}=\frac{n+\alpha}{|B|+K\alpha}
\end{equation} 

\begin{algorithm}
\caption{Gibbs sampler for BTM} 
\textbf{Input}: the number of topics; $K$, hyper-parameters; $\alpha$ ,$\beta$, biterm set; $\mathbf{B}$ 

\textbf{Output}: distribution ; $\boldsymbol\phi_{w|z}$ and $\boldsymbol\theta_z$

	\begin{algorithmic}[1]
	\State initialize topic assignments randomly for all the biterms
		\For {iteration$=1,2,\ldots,N$}
			\For {$b=1,2,\ldots,\mathbf{B}$}
				\State sample $z_{b}$ from $p(z|z_{-b},B,\alpha,\beta)$ 
				\State update statistics $n_{w|z}$ and $n_z$ 
				\State compute the parameter $\boldsymbol\phi_{w|z}$ and $\mathbf\theta_z$
			\EndFor
		\EndFor
	\end{algorithmic} 
\end{algorithm}

\clearpage
\section{Distance Plot of ${\bf A}_k$}
\begin{figure}[!htbp]
\setlength{\fboxsep}{0pt}%
\setlength{\fboxrule}{0pt}%
\begin{center}
\includegraphics[width = 1.0\textwidth]{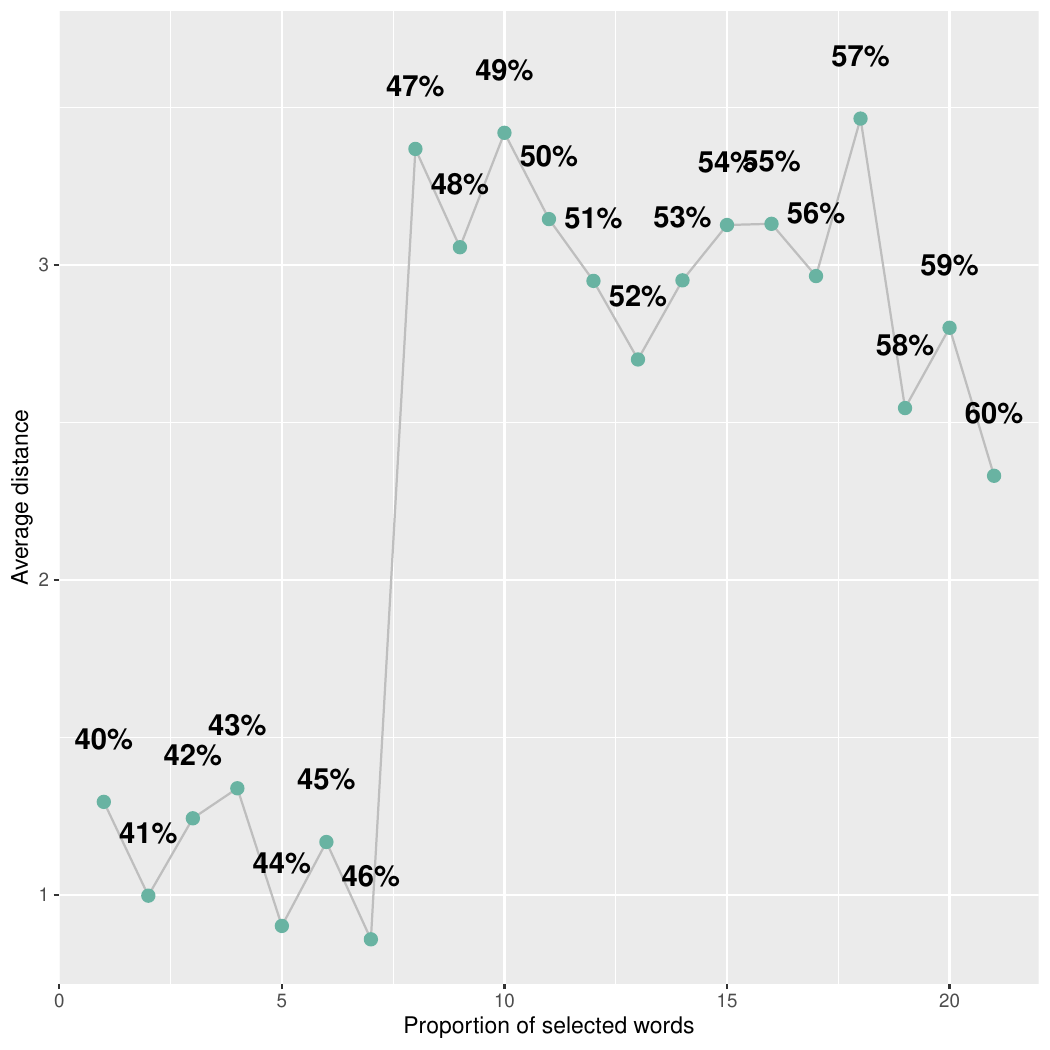}
\end{center}
\caption[]{Distance plot of ${\bf A}_k$}
\end{figure}

\clearpage
\section{ $\textit{Score}$ and Affinity Plots of Topics}

\begin{figure}[!htbp]
\setlength{\fboxsep}{0pt}%
\setlength{\fboxrule}{0pt}%
\begin{center}
\includegraphics[width = 1.0\textwidth]{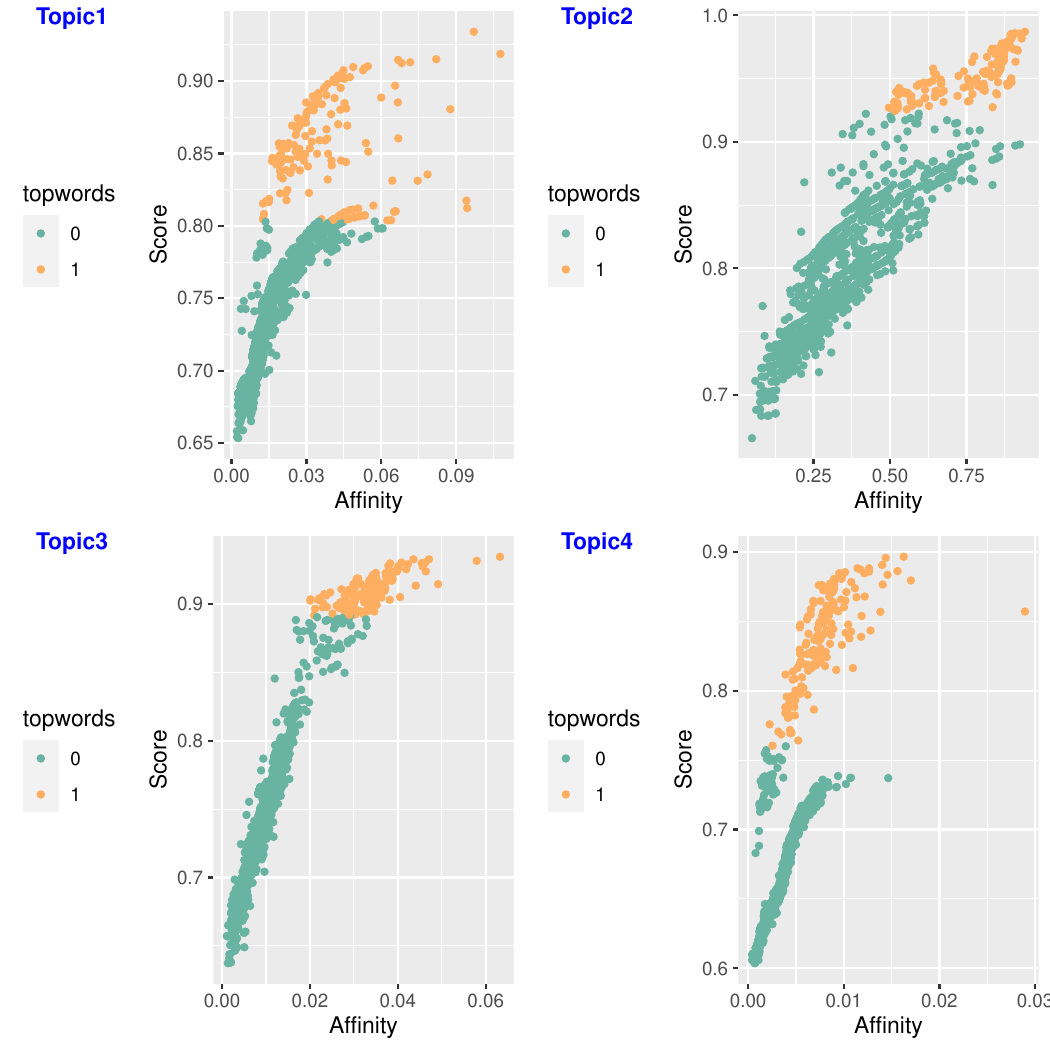}
\end{center}
\caption[]{$\textit{Score}$ and affinity plots of Topics 1 - 4.}
\end{figure}

\begin{figure}[!htbp]
\setlength{\fboxsep}{0pt}%
\setlength{\fboxrule}{0pt}%
\begin{center}
\includegraphics[width = 1.0\textwidth]{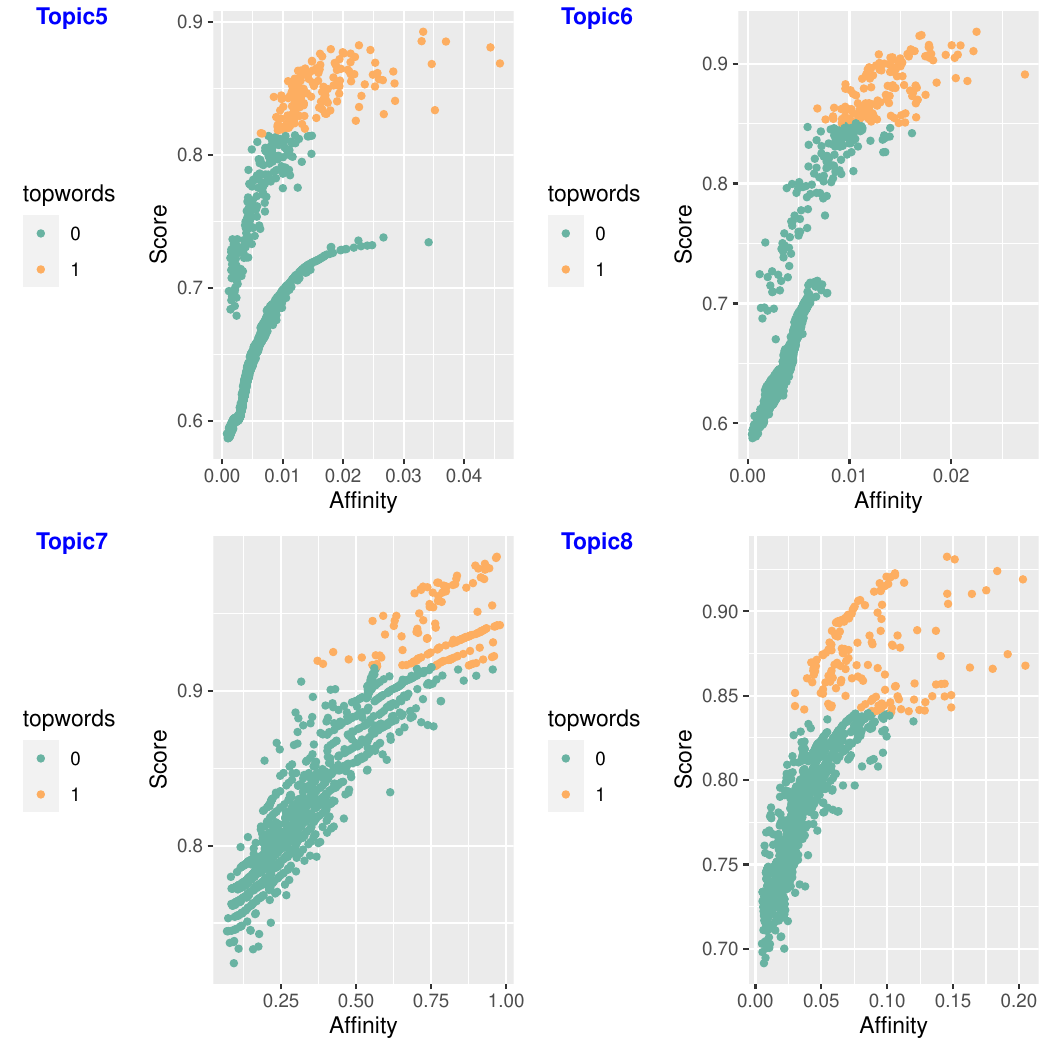}
\end{center}
\caption[]{$\textit{Score}$ and affinity plots of Topics 5 - 8.}
\end{figure}

\begin{figure}[!htbp]
\setlength{\fboxsep}{0pt}%
\setlength{\fboxrule}{0pt}%
\begin{center}
\includegraphics[width = 1.0\textwidth]{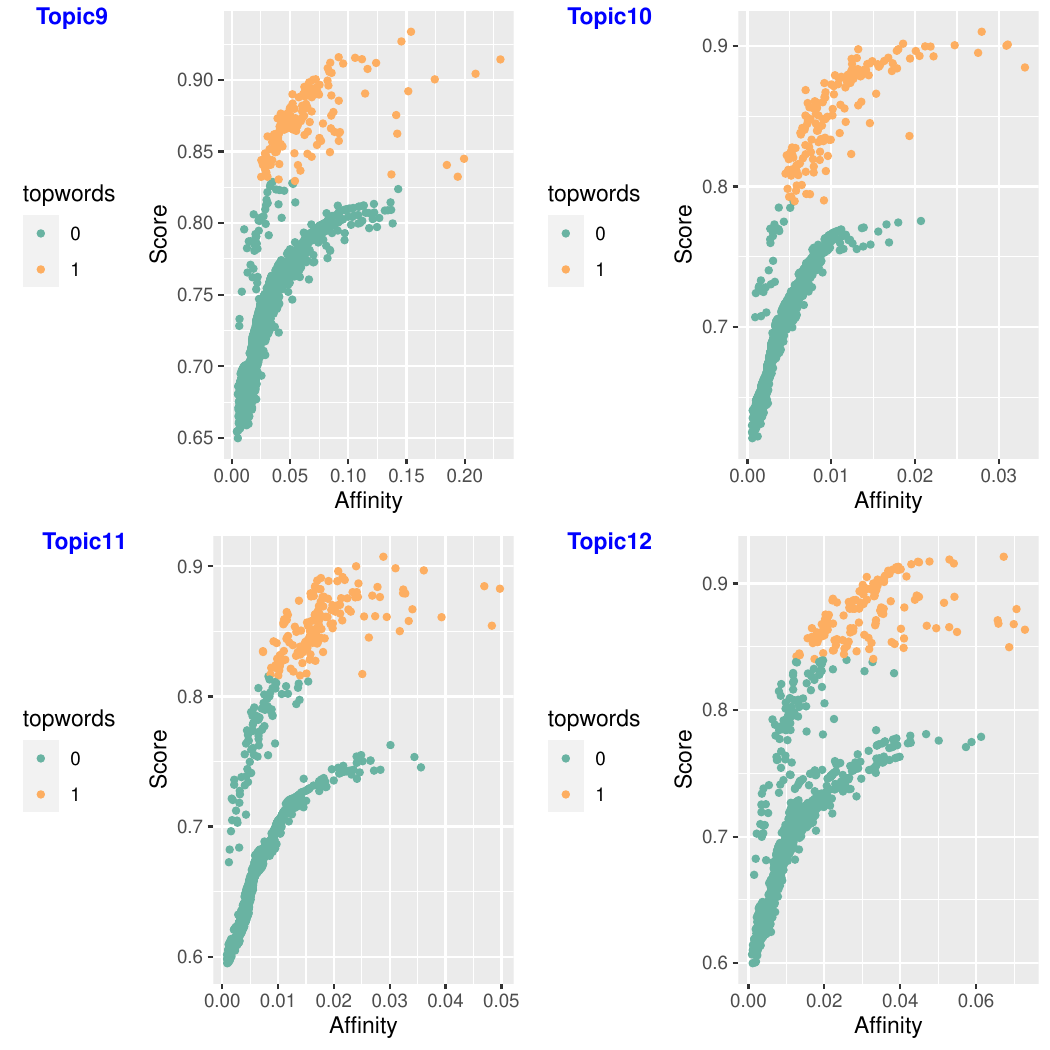}
\end{center}
\caption[]{$\textit{Score}$ and affinity plots of Topics 9 - 12.}
\end{figure}

\begin{figure}[!htbp]
\setlength{\fboxsep}{0pt}%
\setlength{\fboxrule}{0pt}%
\begin{center}
\includegraphics[width = 1.0\textwidth]{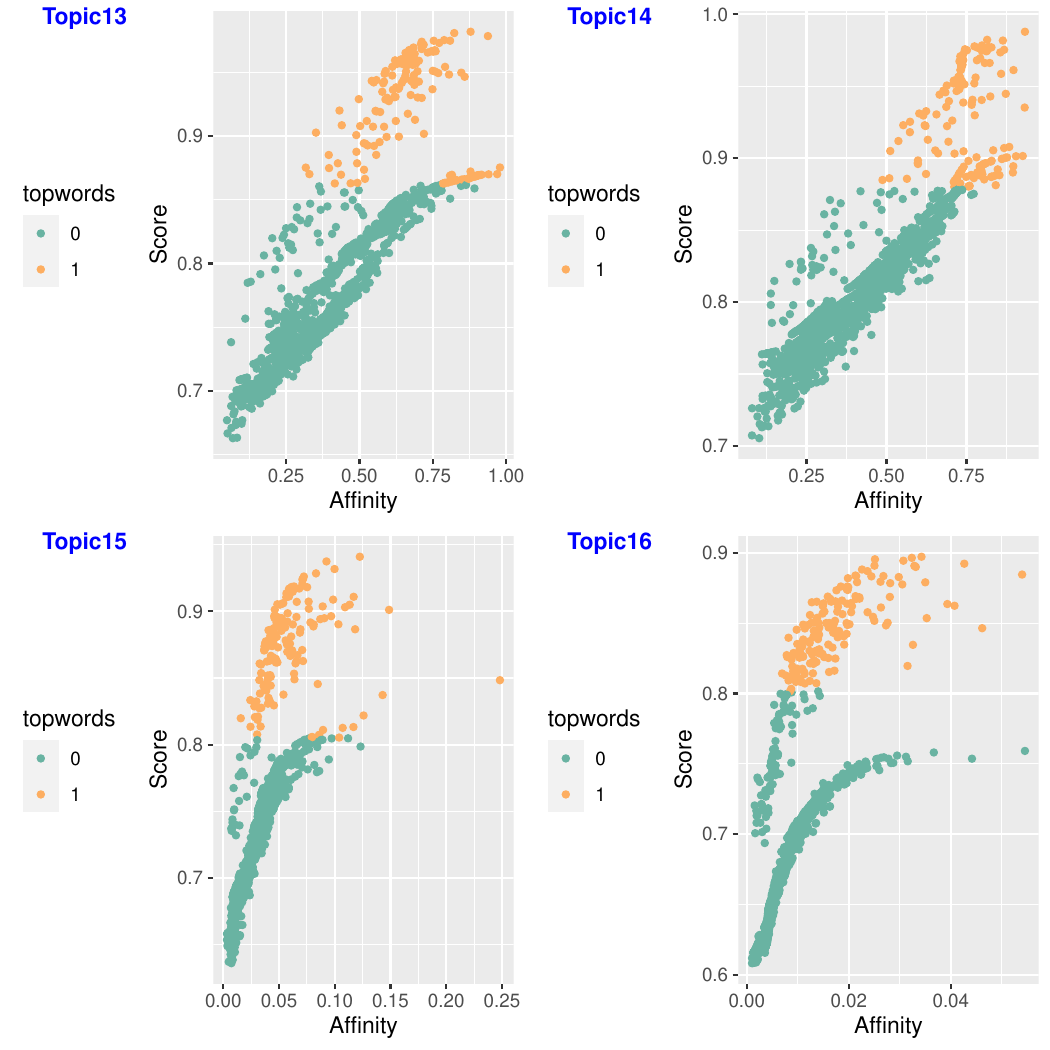}
\end{center}
\caption[]{$\textit{Score}$ and affinity plots of Topics 13 - 16.}
\end{figure}

\begin{figure}[!htbp]
\setlength{\fboxsep}{0pt}%
\setlength{\fboxrule}{0pt}%
\begin{center}
\includegraphics[width = 1.0\textwidth]{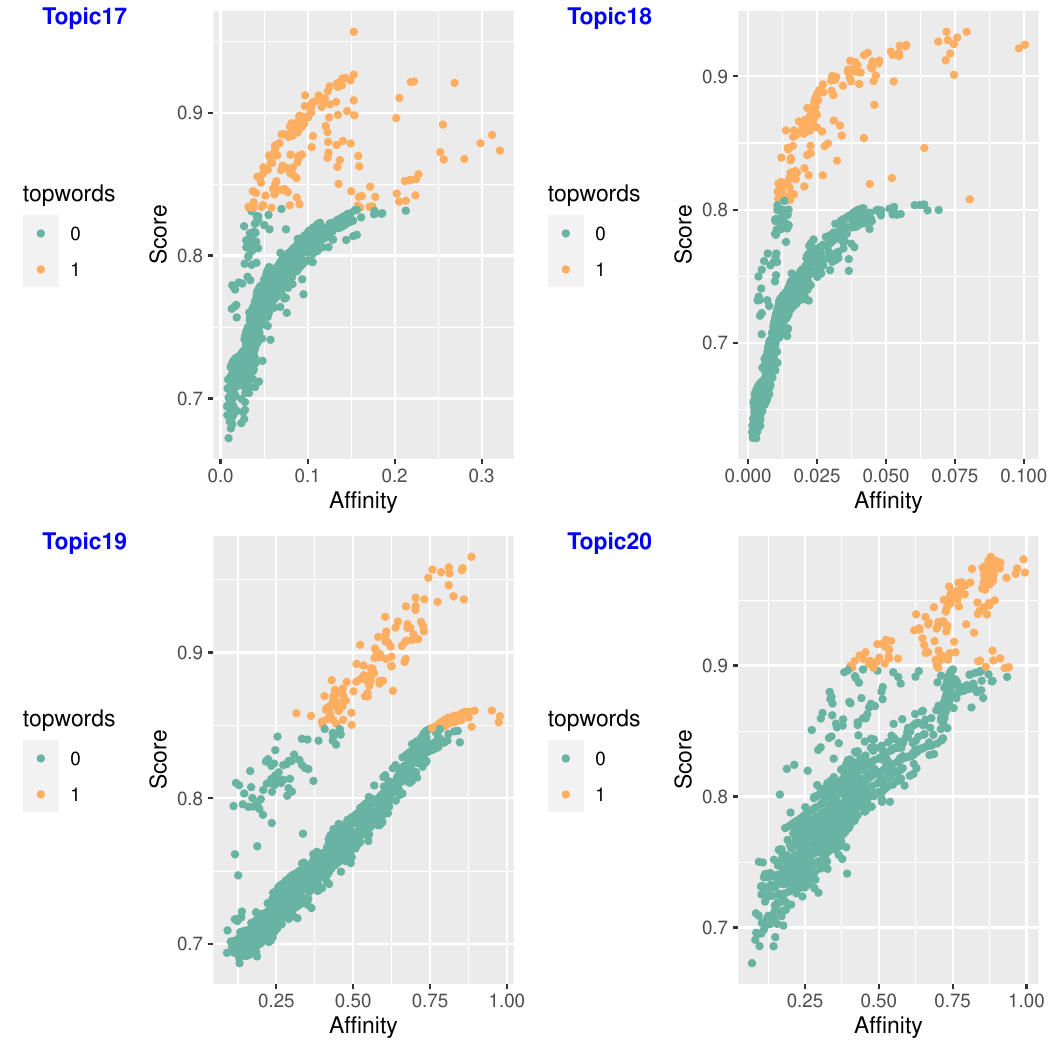}
\end{center}
\caption[]{$\textit{Score}$ and affinity plots of Topics 17 - 20.}
\end{figure}

\clearpage

\section{Topic Interpretation}

\begin{table}[htt]
\centering
\caption{\label{topicname} \textbf{Interpretation of the topic based on $\mathbf{A}^{*}_{47\%}$ matrix.} The abbreviation is marked by an asterisk and the full form can be found in the Web Appendix G.}
\resizebox{\columnwidth}{!}{%
\begin{tabular}{lllll}
  \hline
Topic  & Interpretation & Top score words (score)  \\
  \hline
1 & Lung Scan & subpleural  & crazy paving   & bronchogram\\ 
 & & (0.963) & (0.957) & (0.950)\\
2 & Compound and Drug & Protein Data Bank &  papain-like & intermolecular \\
& & (0.938) & (0.936) &  (0.934) \\ 
3 & Bacteria, Nano, and Diet & $\text{DHA}^{\ast}$ & biofilm  & vancomycin-resistant  \\
&& (0.968) & (0.968) & (0.967) \\ 
4 & Treatment of Other Diseases & sarcoma &  arthroplasty & neoadjuvant \\
&& (0.948) & (0.944) & (0.943)  \\
5 & Symptoms (Cardiovascular) & vein & antithrombotic & $\text{VTE}^{\ast}$  \\
&& (0.906) & (0.899) & (0.898)  \\ 
6 & Molecular-level Response to Infection & $\text{NLRP3}^{\ast}$ & metalloproteinase & upregulation  \\
&& (0.929) &(0.927) &(0.926) \\ 
7 & COVID-19 Risk Prediction Markers & alanin & $\text{BUN}^{\ast}$ & prealbumin \\
&& (0.976) & (0.972) &(0.969)\\ 
8 & Literature Review & Prospero DB &  Wanfang DB  & meta-analys  \\
 && (0.967) & (0.966) & (0.944)\\
9 & Symptom and Comorbidity & petechiae & guillain-barre syndrome & ageusia\\
&&  (0.958) &  (0.955) & (0.952) \\ 
10 & Cardiovascular & infarction & thromboprophylaxis & $\text{VTE}^{\ast}$ \\
&& (0.912) &(0.909) &  (0.909) \\  
11 & Social Impact & classroom  & pedagogy& zoom  \\
&&  (0.925) & (0.911)  &(0.906) \\ 
12 & Financial and Economical Impact & macroeconomics & monetary & profit \\
&& (0.929) &(0.928) & (0.928) \\ 
13 & Statistical Modeling & Weibull& least-square & $\text{MAE}^{\ast}$\\ 
&& (0.984) & (0.975) & (0.966) \\
14 & COVID-19 Test & $\text{LFIAs}^{\ast}$  & transcription-$\text{PCR}^{\ast}$ & Wantai \\
&&  (0.979) & (0.974) &(0.972) \\ 
15 & Psychological/ Mental Issues & anxious&$\text{PTSD}^{\ast}$& post-traumatic \\
&& (0.942) & (0.940) &  (0.935)\\
16 & Lung Scan &  procalcitonin & ground-glass & opacity \\
&& (0.924) &  (0.922) &(0.916) \\
17 & Treatment & reintubation &  multi-centric& Hydroxychloroquine \\
&& (0.943) & (0.943) &  (0.937) \\
18 & Air, Mask, Breathing & laryngoscope& facepiece & supraglottic \\
&& (0.978) & (0.978) & (0.975)\\ 
19 & Prevention of COVID-19 & alcohol-based & decontamination &rub\\
&& (0.944) &  (0.941) & (0.919)\\ 
20 & Immunology and Virus &  $\text{MHC}^{\ast}$ &  multi-epitope & immunodominant \\
&& (0.976) & (0.975) &(0.973)  \\ 
\hline
\end{tabular}%
}
\end{table}

\clearpage
\section{Full Form of Abbreviation}

\begin{table}[h]
\begin{center}
\caption{Abbreviation in Table S2.}
\begin{tabular}{@{}lllll@{}}
\hline
\multicolumn{1}{c}{Abbreviation} & \multicolumn{1}{c}{Full name  }\\ \hline
DHA$^{\ast}$          & Docosahexaenoic   Acid                             &  &  &  \\
VTE$^{\ast}$          & Venous Thromboembolism                             &  &  &  \\
BUN$^{\ast}$          & Blood Urea Nitrogen                                &  &  &  \\
MAE$^{\ast}$          & Mean Bbsolute Error                                &  &  &  \\
LFIAs$^{\ast}$        & Lateral Flow   Immunoassays                        &  &  &  \\
PCR$^{\ast}$          & Polymerase Chain Reaction                          &  &  &  \\
PTSD$^{\ast}$         & Post-traumatic Stress Disorder                      &  &  &  \\
MHC$^{\ast}$          & Major Histocompatibility Complex                   &  &  &  
\\ \hline
\end{tabular}
\end{center}
\end{table}

\bibliography{Reference}